\documentclass[runningheads]{llncs}
\usepackage[T1]{fontenc}
\usepackage{graphicx}
\usepackage{amsmath}
\usepackage{amssymb}
\usepackage{tabularx}
\usepackage{booktabs}
\usepackage{algorithm}
\usepackage[noend]{algpseudocode}
\usepackage{hyperref}
\usepackage{xcolor}
\usepackage{adjustbox}
\usepackage{multirow}
\usepackage{colortbl}

\newcommand{\nfbase}{NonFiction$_{base}$}
\newcommand{\nfrlm}{NonFiction$_{RLM}$}
\newcommand{\frlm}{Fiction$_{RLM}$}
\newcommand{\fnfrlm}{Fiction-NonFiction$_{RLM}$}
\newcommand{\hlt}[1]{{\color{red} #1}}
\renewcommand{\hlt}[1]{\textnormal{#1}}

\begin{document}

\title{Cultural Analytics for Good: Building Inclusive Evaluation Frameworks for Historical IR}

% From Stories to Search: Knowledge Transfer Across Literary Boundaries for Better IR Evaluation
% ⁠From Fiction to Facts: Transfer Learning for Culturally-Aware Search
% Decolonizing Search: Equitable Knowledge Access Through Literary Cross-Training

\titlerunning{Cultural Analytics for Good}

\author{
Suchana Datta
\inst{1}
\orcidID{0000-0001-9220-6652}
\and
Dwaipayan Roy
\inst{2}
\orcidID{0000-0002-5962-5983}
\and
Derek Greene
\inst{1}
\orcidID{0000-0001-8065-5418}
\and
Gerardine Meaney
\inst{3}
\orcidID{0000-0002-5412-5007
} 
\and
Karen Wade
\inst{3}
\orcidID{0000-0002-1547-6672}
\and
Philipp Mayr
\inst{4}
\orcidID{0000-0002-6656-1658}
}
\authorrunning{Datta et al.}

%
%
% \institute{Anonymous Institute}
\institute{
	School of Computer Science, University College Dublin, Ireland \and
	Indian Institute of Science Education and Research, Kolkata, India \and
    School of English, Drama and Film, University College Dublin, Ireland \and
    GESIS -- Leibniz Institute for the Social Sciences, Cologne, Germany\\
	\email{suchana.datta@ucd.ie},
    \email{dwaipayan.roy@iiserkol.ac.in},
	\email{derek.greene@ucd.ie},
    \email{gerardine.meaney@ucd.ie},
    \email{karen.wade@ucd.ie},    \email{philipp.mayr@gesis.org}
}

\maketitle              % typeset the header of the contribution
\begin{abstract}
This work bridges the fields of information retrieval and cultural analytics to support equitable access to historical knowledge. 
Using the British Library’s BL19 digital collection (more than $35,000$ works from $1700-1899$), we construct a benchmark for studying changes in language, terminology and retrieval in the 19th-century fiction and non-fiction. 
Our approach combines expert-driven query design, paragraph-level relevance annotation, and Large Language Model (LLM) assistance to create a scalable evaluation framework grounded in human expertise. 
We focus on knowledge transfer from fiction to non-fiction, investigating how narrative understanding and semantic richness in fiction can improve retrieval for scholarly and factual materials. 
This interdisciplinary framework not only improves retrieval accuracy but also fosters interpretability, transparency, and cultural inclusivity in digital archives. 
% By integrating IR methodologies with humanities-informed perspectives, our work advances the goals of IR4Good - creating evaluation resources and retrieval systems that support richer, historically aware engagement with cultural heritage.
Our work provides both practical evaluation resources and a methodological paradigm for developing retrieval systems that support richer, historically aware engagement with digital archives, ultimately working towards more emancipatory knowledge infrastructures.

\end{abstract}

\section{Introduction} \label{sec:intro}

The digitization of cultural heritage collections has transformed access to historical materials, opening unprecedented opportunities for large-scale cultural and linguistic research. 
Libraries, museums, and archives now host millions of digitized texts spanning centuries of human expression. 
Yet, despite these advances, retrieving and interpreting historical documents remains challenging. 
Variations in language use, evolving genres, and scarcity of annotated data hinder effective information retrieval (IR) and research in this domain. 
For scholars in the humanities, 
%particularly those working on 19th-century texts, 
existing retrieval systems often fall short of supporting nuanced, interpretive, and context-sensitive exploration.

At the same time, information retrieval research increasingly emphasizes social good, with growing attention to fairness, transparency, and equitable access to knowledge. 
The IR for Good initiative seeks to develop methods that extend beyond accuracy, addressing the ethical and epistemic dimensions of information access. 
In the context of cultural heritage, this means designing retrieval frameworks that not only perform well technically, but also promote inclusivity and interpretability across diverse historical 
%voices and 
genres.
Our work contributes to this agenda by bridging information retrieval and cultural analytics to support equitable engagement with 19th-century British literature and non-fiction. 
Using the British Library Digital Collection (BL19)\footnote{\scriptsize See the British Library's Digital Collections: \url{https://labs.biblios.tech}} of over $35,000$ digitized works published between $1700$ and $1899$, we develop a benchmark for studying language change and cross-genre retrieval. 
We frame the 19th-century as a dynamic period of linguistic, social, and cultural transformation, offering an ideal testbed for examining how retrieval systems handle evolving semantics and stylistic variation.
To address the dual challenges of scalability and domain expertise, we combine expert-driven query design, LLM-assisted relevance annotation, and expert validation. 
This hybrid approach enables the creation of reliable, fine-grained relevance judgments, 
%at the paragraph level, 
capturing the interpretive depth required for cultural-analytic research while maintaining reproducibility and coverage. 

Furthermore, we investigate the applicability of knowledge transfer between fiction and non-fiction retrieval, exploring whether the rich narrative structures and semantic complexity of fiction can support more effective retrieval in factual contexts.
The motivation behind this framework arises from the way people often learn and recall information across different narrative forms. 
Fiction frequently serves as an interpretive bridge to reality, presenting real events, people, and social conditions through imaginative storytelling.
For instance, a reader may first encounter a real historical event, such as a war, an invention, or a political movement, through a fictional story that blends factual detail with narrative framing.
Later on, the same reader might search for information about that real event, with their queries shaped by what they learned from the fiction, i.e. narrative cues, character references, or metaphorical descriptions that do not always match the vocabulary of factual records. 
This idea connects with cognitive studies suggesting that fiction supports semantic generalization and knowledge integration across domains~\cite{MARSH2003519,jacobs2015neurocognitive,mar2018stories}. 

From an IR perspective, such cross-genre information seeking remains underexplored, though related phenomena have been observed in domain adaptation~\cite{bendavid2010theory} and semantic transfer~\cite{diaz2016local,zamani2018neural}, where representations learned in one context improve retrieval in another. 
By testing whether term associations derived from fiction can enhance the retrieval of non-fictional materials, we aim to simulate and support this natural, human process of connecting narrative understanding with factual knowledge.
\hlt{Precisely, this paper focuses on the use case of \emph{non-fiction search}, testing whether narrative-informed relevance models derived from \emph{fiction} can enhance retrieval performance and interpretability in factual, scholarly contexts.}

% ========= add
Besides, this work directly addresses the concerns about social good and bias by engaging with the structural inequalities embedded in BL19. 
As a collection shaped by English language and imperial knowledge production, BL19 reflects asymmetries in who has historically been able to access, interpret, and fund archival research. 
Scholars in former colonies, particularly in Africa and parts of Asia, often face significant barriers to consulting materials about their own histories, which are held in European institutions, such as the British Library. 
By making the BL’s historical, topographical and travel collections more easily accessible in digital form, our work lowers barriers to entry for researchers, students, and educators globally, especially those in under-resourced contexts.
While this does not remove the collection’s inherent bias, it constitutes a concrete intervention that improves equity of access and enables more diverse and critical historical scholarship.
This is particularly pressing in the context of the emergence of global 19th century studies. 
As Valdez argues: ``A global orientation is not inherently anti-imperialist nor anti-racist; purported global approaches often retain a Eurocentric core, expanding the scope of geographical coverage while centring the categories of empire and nation" \cite{valdez}.  
In improving access and facilitating analysis of an imperial archive for scholars globally, this paper contributes to socially beneficial information access by advancing equitable and interpretable retrieval. 
By developing a benchmark grounded in the 19th-century British Library collections, we enable transparent and reproducible research that will support scholars, educators, and the public in critically engaging with an important repository of historical data and cultural heritage.

% \hlt{
% Specifically, this paper contributes to socially beneficial information access by advancing equitable and interpretable retrieval for cultural and historical materials. 
% By developing a benchmark grounded in the 19th-century British Library collections, we enable transparent and reproducible research
% % evaluation of retrieval systems 
% that will support scholars, educators, and the public in engaging with cultural heritage.
Specifically, our hybrid relevance framework, combining expert insight with scalable, LLM-assisted annotation, lowers the barrier to creating high-quality historical datasets across underrepresented domains.
Furthermore, our cross-genre analysis of knowledge transfer between fiction and non-fiction demonstrates how computational approaches to narrative and factual knowledge can bridge literary and scholarly inquiry, thereby fostering richer access to cultural memory. 
%In doing so, this work aligns with broader social goals of democratizing access to historical digital resources, promoting inclusive scholarly works, and ensuring that advances in retrieval technology serve both academic and public understanding of the past.
% }
In summary, this paper makes the following key contributions.
\begin{itemize}
    \item \textbf{A novel, expert-grounded benchmark for historical retrieval.}
    We present a curated evaluation dataset from the 19th-century British Library collection with $35$ expert-informed queries spanning fiction and non-fiction. 
    The dataset is constructed using a scalable hybrid pipeline combining LLM-assisted graded relevance judgments with expert verification to ensure high agreement, reproducibility, and historical validity.
    
    % \item \textbf{A novel benchmark for historical retrieval.} 
    % We introduce a curated evaluation dataset based on the British Library’s 19th-century collection, featuring $35$ expert-informed queries spanning fiction and non-fiction contexts of 19th-century British writings.
    % \item \textbf{A scalable, expert-grounded relevance framework.} We develop a hybrid annotation pipeline that integrates LLM-assisted graded relevance judgments with human expert verification, achieving near-perfect agreement while maintaining scalability and reproducibility.
    \item \textbf{A cross-genre retrieval framework based on knowledge transfer.} We investigate knowledge transfer between fiction and non-fiction, demonstrating how narrative understanding and semantic richness in fiction can enhance retrieval performance for factual and scholarly materials.
\end{itemize}
Our curated British Library collection, 
of fiction and non-fiction texts, 
the set of queries, and relevance judgments, together with the code used for our study, are publicly available to ensure full reproducibility and to support further research on cross-genre information retrieval\footnote{\url{https://github.com/suchanadatta/BL19-benchmark-knowledge-transfer.git}}.

% \textbf{PLACE IT SOMEWHERE}:

\section{Related Work} \label{sec:rel-work}

\paragraph{\textbf{Historical IR.}}
Historical and cultural collections pose unique challenges for information retrieval due to diachronic language variation, genre diversity, and limited ground-truth annotations.
Prior work in this area has focused on improving retrieval and representation of historical texts through techniques, such as temporal language modeling \cite{temporal}, domain adaptation for historical corpora \cite{adaptation}, semantic normalization for spelling and vocabulary shifts \cite{vocab_shift}, and evolution of term usage over the decades \cite{datta2025ecir,datta2025jcdl}. 
However, most historical IR studies rely on small-scale evaluation datasets or heuristic relevance signals, limiting reproducibility and cross-domain generalization. 
Our work extends this line of research by introducing a benchmark collection for 19th-century British writings with graded relevance judgments, enabling systematic evaluation of retrieval models in a historically grounded setting.

\paragraph{\textbf{Transfer learning from fiction to non-fiction.}}
Fictional texts (books, stories) are large, coherent, and rich in event sequences, character relations and causal chains. 
Language models that learn from such content can acquire broad lexical, syntactic, discourse and narrative patterns that plausibly benefit downstream tasks even when those targets are non-fictional (e.g., question-answering over encyclopedic text, inference about events described in news, or co-reference in technical prose). 
This motivation underpins much practical pre-training, early and influential language models used large fiction collections as part of their unsupervised pre-training corpora \cite{bert}.

BookCorpus \cite{bookcorpus}, a large collection of books across genres (includes romance, fantasy, and sci-fi), has been widely used as part of pre-training corpora for language models.
Notably, Google’s BERT \cite{bert} and subsequent optimization efforts (e.g. RoBERTa \cite{roberta}) included BookCorpus in their pre-training data.
These models are pre-trained on combinations of books and web or Wikipedia text and achieved strong performance on many downstream, often non-fictional, benchmarks (GLUE~\cite{wang2018glue}, SQuAD~\cite{rajpurkar2018squad}), demonstrating effective transfer from mixed fiction sources to non-fiction. 
These results show that representations learned from fiction-enriched corpora can be useful for factual and reasoning tasks after fine-tuning; however, the pre-training corpora used in practice are a mixture of books, Wikipedia, and news.
Therefore, the literature generally shows transfer from mixed corpora that include fiction, rather than isolated studies proving fiction works better than matched non-fiction corpora \cite{transfer1,transfer2}.

Research in cognitive psychology and cultural studies suggests that fiction can support information retrieval in non-fiction domains. 
Readers often acquire factual knowledge from narratives, and fictional contexts shape conceptual associations and term usage relevant to search behavior. 
Marsh et al. \cite{MARSH2003519} show experimentally that people learn facts from fictional texts, providing a psychological basis for cross-genre knowledge transfer. 
Similarly, cultural studies note that literary works can transmit domain knowledge (e.g., scientists writing fact-based novels), positioning fiction as a vehicle for informal knowledge exchange and a potential source of useful semantic cues for non-fiction retrieval \cite{Azagra_Caro_2018}.
While several lines of research motivate using fiction as a source for improving retrieval on non-fiction, cognitive and educational research shows that readers acquire factual knowledge from narrative sources, and that fiction can shape conceptual associations that affect how queries map to information needs \cite{MARSH2003519}. 
% For instance, Marsh et al. \cite{MARSH2003519} experimentally demonstrate that people learn and integrate facts from fictional texts under many conditions, establishing a psychological basis for cross-genre knowledge transfer.
% 
% Work in cultural studies has also documented how literary texts can contain and transmit domain knowledge \cite{Azagra_Caro_2018} (e.g., scientists writing novels that embed factual content), framing literary fiction as a potential vehicle for informal knowledge transfer. 
% This supports the idea that fiction sometimes encodes factual or semantic cues that could be useful when searching in non-fiction collections.

From an NLP or IR standpoint, the problem reduces to domain or genre shift. 
Models trained on one genre may learn lexical, stylistic, and conceptual patterns that are partially useful for another \cite{jurczyk-choi-2017-cross}, if those patterns capture semantic relations or paraphrases that help match queries to relevant documents. 
Such cross-genre IR tasks (formal to conversational, news to social, etc.) have been formulated and show that this is tractable with the right methods.

Research has shown that models trained on narrative tasks learn representations of events, causality and character interactions that improve performance on certain common sense inference and story-reasoning tasks. 
However, direct, systematic demonstrations that such narrative-trained representations improve purely non-fiction tasks (e.g., encyclopedia-style question answering, information extraction from scientific articles) are comparatively scarce \cite{mostafazadeh-etal-2016-corpus}; most work demonstrates gains on other narrative or common sense benchmarks.

Although prior work provides a strong methodological foundation for transfer learning from fiction to its non-fiction counterpart, there are limited works that (i) construct a large-scale benchmark dataset for historical fiction and non-fictional documents, and (ii) systematically evaluate cross-genre knowledge transfer from fiction to factual retrieval tasks.
Our contributions in this paper address these gaps by creating a reproducible benchmark together with an empirical evaluation framework, linking information retrieval, computational literary studies, and archival findability.

\section{Dataset and Benchmark Development}

The design of any IR evaluation typically follows the Cranfield paradigm~\cite{cleverdon}, which includes three essential components: a document collection, a set of information needs in the form of queries, and relevance judgments that link the former two components.
Building on this paradigm, our goal is to create a benchmark that 
% not only adheres to this structure, but also 
reflects the linguistic, stylistic, and semantic diversity of 19th-century writing.
In this section, we describe the construction of the benchmark dataset.

\subsection{Corpus}
In our experiments, we analyze two distinct subsets of the BL19 collection, classified according to the metadata provided by the British Library, which describes their holdings from the 19th-century. 
The first subset consists of $10,210$ English language works of fiction, published between $1830$ and $1899$.
This subset features well-known novels by authors such as Charles Dickens and Jane Austen, as well as many lesser-known writers. 
The second subset comprises $15,780$ English language non-fiction books from the same period, encompassing a wide range of subjects, including history, geography, philosophy, and travel that reflect the thematic diversity of 19th-century publications held by the British Library.

\subsection{Topics}

The selection of topics followed a structured, expert-informed process to ensure domain relevance and diversity. 
We collaborated with specialists in cultural analytics and 19th-century British literature to identify queries reflecting core research themes and interpretive challenges. 
Experts proposed candidate topics based on their knowledge of genres, historical issues, and key debates, which were refined to avoid redundancy and ensure broad thematic coverage across social, political, and aesthetic dimensions. 
Through iterative consultation, $35$ potential queries were finalized, representing a balanced mix of fictional and non-fictional contexts. 
This expert-driven design captures the conceptual richness and linguistic variability of humanities-oriented information retrieval tasks while aligning with common scholarly and public search practices.
Considering common search patterns in both scholarly and public search behavior, we kept the queries intentionally with an average query length of $2.8$, primarily keyword-style, to reflect realistic, user-oriented search formulations.

\subsection{Relevance Judgments}

For each of the $35$ expert-curated queries, we retrieved the top $100$ documents using the BM25 \cite{bm25,bm25_beyond} ranking function as a baseline retrieval model.
To ensure a richer and more diverse candidate pool for relevance assessment, we further generated multiple query variants per original query by further consulting the experts, capturing linguistic, syntactic, and contextual variations observed across historical and modern usage.
The resulting query–document pairs were then assessed for relevance using \texttt{gpt-5-mini}, which we employed as an automated assessor.
Relevance was graded on a five-point ordinal scale ($0-4$), where `0' indicates non-relevant and `4' signifies highly relevant. 
The LLM was provided with query–document pairs in context, along with task-specific instructions emphasizing interpretive and cultural relevance within the domain of 19th-century British writings. 
This allowed the model to capture nuanced thematic and stylistic relationships beyond surface lexical overlap.

To ensure reliability, we randomly sampled 40\% of the judgments generated by the LLM and verified them independently by field experts in cultural analytics. 
Inter-assessor agreement between \texttt{gpt-5-mini} and the experts was nearly 100\%, with discrepancies observed only at the extreme ends of the scale -- cases where one assessor assigned a `4' while the other gave a `3', or one marked a `0' while the other gave a `1'.
In such instances, the final decision was made by the expert assessor, confirming the high consistency of automated relevance estimation in this specialized domain.
% This hybrid evaluation protocol, combining automated grading with expert validation, balances scalability and domain sensitivity, providing a robust and reproducible framework for evaluating retrieval performance in humanities-oriented IR tasks.

% \subsection{Retrieval Performance} -- ? Move later?

\section{Evaluation}

For evaluation, we employed two primary retrieval models: (i) the BM25 \cite{bm25,bm25_beyond} ranking function, a probabilistic retrieval model that serves as a strong lexical baseline based on term frequency saturation and document length normalization, and (ii) the Relevance-based Language Model (RLM) \cite{DBLP:conf/trec/JaleelACDLLSW04,10.1145/383952.383972}, which represents the class of pseudo-relevance feedback (PRF) and query expansion techniques. 
% RLM estimates a relevance model from the top-ranked documents of an initial retrieval, expanding the query with high-probability terms to improve recall and capture latent semantic associations within the corpus.

Note that we intentionally do not employ neural re-rankers in this study.
Our goal is to isolate and analyze the effects of cross-genre knowledge transfer, specifically how narrative structures, learned from fiction, can enhance retrieval in non-fiction -- without introducing confounding factors from large, pre-trained language models optimized on modern text.
Neural re-rankers, such as BERT-based models~\cite{nogueira2019passage,santhanam2022colbertv2}, tend to rely heavily on contemporary linguistic patterns and dense semantic encodings, which may not align well with the stylistic and lexical characteristics of 19th-century English texts.
Moreover, due to the scarcity of supervised relevance labels in historical corpora, fine-tuning such models would be unreliable and risk overfitting.
By focusing on lexical and probabilistic feedback models, we maintain interpretability, transparency, and historical validity. This ensures that any observed improvements truly stem from genre-informed knowledge transfer rather than from opaque neural scoring mechanisms.

However, integrating neural re-ranking represents a promising avenue for future research.
Genre-adapted dense retrievers or hybrid models fine-tuned on historical corpora~\cite{meng2024aug,ance} could complement the interpretability of probabilistic models with the representational depth of neural architectures.
Such extensions would allow deeper exploration of how narrative semantics captured by large language models can be aligned with historically grounded retrieval tasks.

\subsection{Baseline Models Evaluated}

\begin{itemize}
    \item {\textbf{\nfbase}} -- 
    \hlt{This configuration defines the non-fiction baseline, applying the BM25 \cite{bm25,bm25_beyond} ranking model to the BL19 non-fiction subset. 
    It provides a reference for retrieval effectiveness using a standard lexical approach on historical texts with factual, expository language.}
    % This configuration represents the non-fiction baseline, where the BM25 \cite{bm25,bm25_beyond} ranking model is applied to the non-fiction subset of the BL19 collection. 
    % It establishes a reference point for retrieval effectiveness using a standard lexical matching approach.
    % This baseline helps quantify how well traditional retrieval methods perform on historical non-fiction texts characterized by factual, expository language.

    \item {\textbf{\nfrlm}} --
    \hlt{In this setup, we apply RM3-based pseudo-relevance feedback~\cite{DBLP:conf/trec/JaleelACDLLSW04,10.1145/383952.383972} on the non-fiction collection, tuning parameters on a validation subset for optimal graded relevance performance. 
    \nfrlm~assesses how intra-domain relevance modeling enhances non-fiction retrieval by capturing domain-specific term associations without cross-genre influence.}
    % In this setup, we apply pseudo-relevance feedback (PRF) using the RM3~\cite{DBLP:conf/trec/JaleelACDLLSW04,10.1145/383952.383972} model on the non-fiction collection.
    % Model parameters were tuned on a held-out validation subset to optimize retrieval performance under graded relevance evaluation. 
    % The goal of~\nfrlm~is to measure the degree to which non-fiction retrieval can be improved through intra-domain relevance modeling, capturing domain-specific term associations without introducing cross-genre signals.
\end{itemize}

% \textbf{PLACE IT PROPERLY}

\subsection{Cross-Genre Knowledge Transfer}
% Motivate why we are establishing our knowledge transfer framework on the query expansion setup.

We base our knowledge transfer framework on a query expansion setup, as it offers a clear and interpretable way to examine how semantic knowledge from fiction can be applied to non-fiction retrieval. 
Query expansion enriches a user’s information need with related terms, making it well-suited for studying how narrative and thematic associations transfer across genres. 
Using pseudo-relevance feedback models, we can trace how specific terms and distributions derived from fiction influence non-fiction retrieval. 
This transparency is crucial in cultural and historical contexts, where interpretability and scholarly accountability are as important as performance gains.
Expansion-based transfer thus reveals how fictional cues, metaphors, archetypes, or emotional lexicons affect factual retrieval, motivating our following two cross-genre evaluation models.

% We establish our knowledge transfer framework on the query expansion setup. 
% This is because it provides a natural and interpretable mechanism for examining how semantic knowledge learned in one domain (fiction) can be transferred and applied to another (non-fiction). 
% Query expansion explicitly models the process of enriching a user’s information need with additional contextual or semantically related terms, making it an ideal testbed for studying how narrative and thematic associations learned from one genre influence retrieval performance in another.

% Pseudo-relevance feedback models, such as RM3, allows us to trace the influence of specific terms and distributions derived from different corpora. 
% This transparency is especially valuable in cultural and historical retrieval, where interpretability and scholarly accountability are as important as performance gains. 
% Through expansion-based transfer, we can observe which lexical or conceptual cues from fiction (e.g., metaphors, social archetypes, or emotional lexicons) enhance or distort retrieval in factual non-fiction contexts.
% Hence, we employ the following two models for cross-genre evaluation.
%
\begin{itemize}
    \item {\textbf{\frlm}} -- 
    This experiment examines cross-genre knowledge transfer from fiction to non-fiction. 
    RLM is trained on the fiction collection to learn narrative and semantic expansion terms, which are then applied to retrieve non-fiction documents. 
    This setup tests whether the semantic richness and narrative coherence of fiction can improve the interpretability and effectiveness of non-fiction retrieval.

    \hlt{
    However, this cross-collection feedback strategy naturally faces a vocabulary mismatch problem -- the terms most salient in fiction may not directly appear in non-fiction.
    To address this, we adapt a term selection strategy proposed in~\cite{carpineto2001info}
    % In this method, 
    where,
    candidate expansion terms are first ranked using one weighting function, and weighted using another, allowing more robust term selection across heterogeneous corpora.
    In our context, we select only those expansion terms that overlap between fiction and non-fiction collections, while assigning weights based on their importance within the fiction paragraphs.
    % This ensures that the expansion model captures the conceptual expressiveness of fiction while remaining lexically aligned with the target non-fiction domain.
    }
    % \cite{roy2019rm3}.\cite{carpineto2001info}
    % This experiment explores cross-genre knowledge transfer from fiction to non-fiction. 
    % The RM3 model is first applied on the fiction collection to learn expansion terms and relevance distributions that reflect narrative and semantic patterns typical of 19th-century fiction. 
    % The resulting relevance model is then used to retrieve documents from the non-fiction collection. 
    % This setup investigates whether representations derived from fiction, characterized by richer narrative structures and broader semantic associations, can enhance retrieval in the factual and scholarly domain. 
    % It directly tests our hypothesis that semantic richness and narrative coherence learned from fiction can improve retrieval interpretability and effectiveness for non-fiction materials.
% 
    \item {\textbf{\fnfrlm}} -- Finally, this configuration combines the fiction and non-fiction subsets into a unified collection. 
    RLM feedback is computed over this joint corpus, allowing the relevance model to incorporate term distributions and thematic signals from both genres. 
    The resulting query expansion thus reflects a genre-agnostic relevance model, balancing factual precision with narrative expressiveness. 
    This setup evaluates whether integrated modeling across genres can yield more generalizable retrieval behavior, improving robustness and interpretive breadth compared to genre-specific or unidirectional transfer approaches.
\end{itemize}

\subsection{Evaluation Metrics and Experimental Setup}

We evaluate retrieval effectiveness using a suite of standard IR metrics that capture both ranking quality and user-oriented performance.
Specifically, we report Mean Average Precision (MAP) as an overall indicator of precision across recall levels, Recall to measure the completeness of relevant item retrieval, and Normalized Discounted Cumulative Gain (nDCG) to account for the graded relevance levels ($0-4$) in our annotations. 
In addition, Precision at rank 10 ($P@10$) reflects the quality of the top results typically viewed by users, while Mean Reciprocal Rank (MRR) evaluates the position of the first relevant document in each ranked list. 
Together, these complementary measures provide a balanced assessment of both early precision and overall retrieval effectiveness across our experimental configurations.

We employ the Lucene\footnote{ \url{https://lucene.apache.org/}} implementation of the probabilistic retrieval model BM25.
For all experiments, the default parameter values of $k_1$ and $b$ are used.
In the RLM experiments, we vary the number of top-ranked documents $M$ in the range of $\{10,20,\dots,100\}$. 
Another key parameter is the number of terms $T$ with the highest weight values obtained from the feedback model.
% These term weights are used to compute Kullback–Leibler (KL) or Jensen–Shannon (JS) divergences for re-ranking in the standard RLM framework \cite{10.1145/383952.383972}.
The parameter $T$ is varied from $20$ to $120$ in increments of $10$.
We report results with the optimal settings of $M$ and $T$ in Table \ref{tab:result}.

% \section{Transferring Knowledge from Fiction to Non-fiction}

% \subsection{Methods}

% \paragraph{\textbf{\nfbase.}} BM25 on non-fiction.

% \paragraph{\textbf{\nfrlm.}} RM3 on non-fiction collection. Tell about the experimental settings.

% \paragraph{\textbf{\frlm.}} RM3 on fiction collection, and then retrieval from the non-fiction.

% \paragraph{\textbf{\fnfrlm.}} RM3 on the combined fiction and non-fiction collection.

\subsection{Results and Analysis}

\begin{table}[!t]
\centering
\caption{
\small
Comparison of retrieval effectiveness between different techniques, \frlm~and a number of baselines. 
Improvements with \frlm~found out to be statistically significant with respect to most of the baselines (t-test with $p<0.05$).  
Superscript $*$ and $\dagger$ respectively indicate significant improvement compared to \nfbase~and \nfrlm.
}
\begin{adjustbox}{width=0.8\textwidth}
\begin{tabular}
{l@{~~~} | @{~~~}c@{~~~}c@{~~~}c @{~~~}c@{~~~}c}

\toprule

Methods & MAP & Recall & nDCG & P@10 & MRR\\

\midrule

\nfbase
& 0.4993 & 0.5942 & 0.5653 & 0.5122 & 0.6071 \\

\nfrlm
& 0.5592$^*$ & 0.6536$^*$ & 0.6059$^*$ & 0.5480 & 0.6375 \\

\frlm
& \textbf{0.5743$^{*\dagger}$} & \textbf{0.6944$^{*\dagger}$} & \textbf{0.6183$^{*\dagger}$} & \textbf{0.5511$^*$} & \textbf{0.6401$^*$}\\

\fnfrlm
& 0.5411$^*$ & 0.6454$^*$ & 0.5989$^*$ & 0.5371 & 0.6276 \\

\bottomrule

\end{tabular}
\label{tab:result}

\end{adjustbox}
\end{table}
Table~\ref{tab:result} presents retrieval performance across four configurations on the BL19 benchmark, evaluated using standard IR metrics.
The \nfbase~configuration (BM25 on non-fiction) provides the lexical baseline.
The MAP of $0.4993$ and nDCG of $0.5653$ confirm that traditional term-matching remains competitive even in this historical corpus.
However, the relatively modest recall ($0.5942$) indicates that purely lexical models under-represent semantically related but lexically divergent passages; this could be due to the orthographic and stylistic variation in historical corpora (usage of old English).
This aligns with earlier findings that vocabulary drift and genre-specific expression may reduce the effectiveness of standard retrieval techniques on diachronic text~\cite{chi2024diachronic,hamilton2016diachronic}.

Introducing RLM in the same non-fiction domain
(\nfrlm) yields consistent improvements across all metrics, over 12\%, 10\%, and 7\% improvement respectively over MAP, recall and nDCG, relative to the baseline.
The \frlm~configuration (i.e., RLM learned from fiction and applied to non-fiction for retrieval) produces the best overall results across nearly all metrics.
This confirms our central hypothesis that semantic and narrative associations learned from fiction can beneficially transfer to factual retrieval.
Fictional prose from the same historical period encodes broader conceptual relationships (e.g., metaphoric or causal associations between `events', `invention', `progress', etc.) that are semantically relevant to non-fiction topics but may be absent in strictly factual co-occurrence patterns.
The approximate improvement of over 3\%  in MAP and 4\% in recall over the intra-domain model (\nfrlm) indicates that narrative-derived expansion terms bridge lexical and stylistic gaps within historical writing.
As evident from the table, the nature of the top-ranked results remains largely unchanged, as reflected by the similar P@10 and MRR values.
The gains are most notable in MAP and recall, suggesting fiction-derived term distributions help the system retrieve additional related passages lower in rank.
This pattern suggests an improvement in coverage and conceptual breadth, rather than in surface-level lexical precision.

When both fiction and non-fiction are merged for relevance modeling (\fnfrlm), performance decreases slightly relative to the fiction-only transfer setting.
Although recall remains higher than the pure BM25 baseline, the mixture of genre-specific signals appears to dilute the discriminative value of feedback terms.
This result shows a trade-off between semantic richness and topical precision.
Fiction adds useful contextual diversity, but mixing very different sources without control can reduce focus in retrieval.
Similar effects have been observed in earlier work on cross-domain feedback and pseudo-relevance transfer \cite{lv2010prlm,diaz2016local}.

\begin{figure*}[!t]
\centering
\includegraphics[width = 1\textwidth]{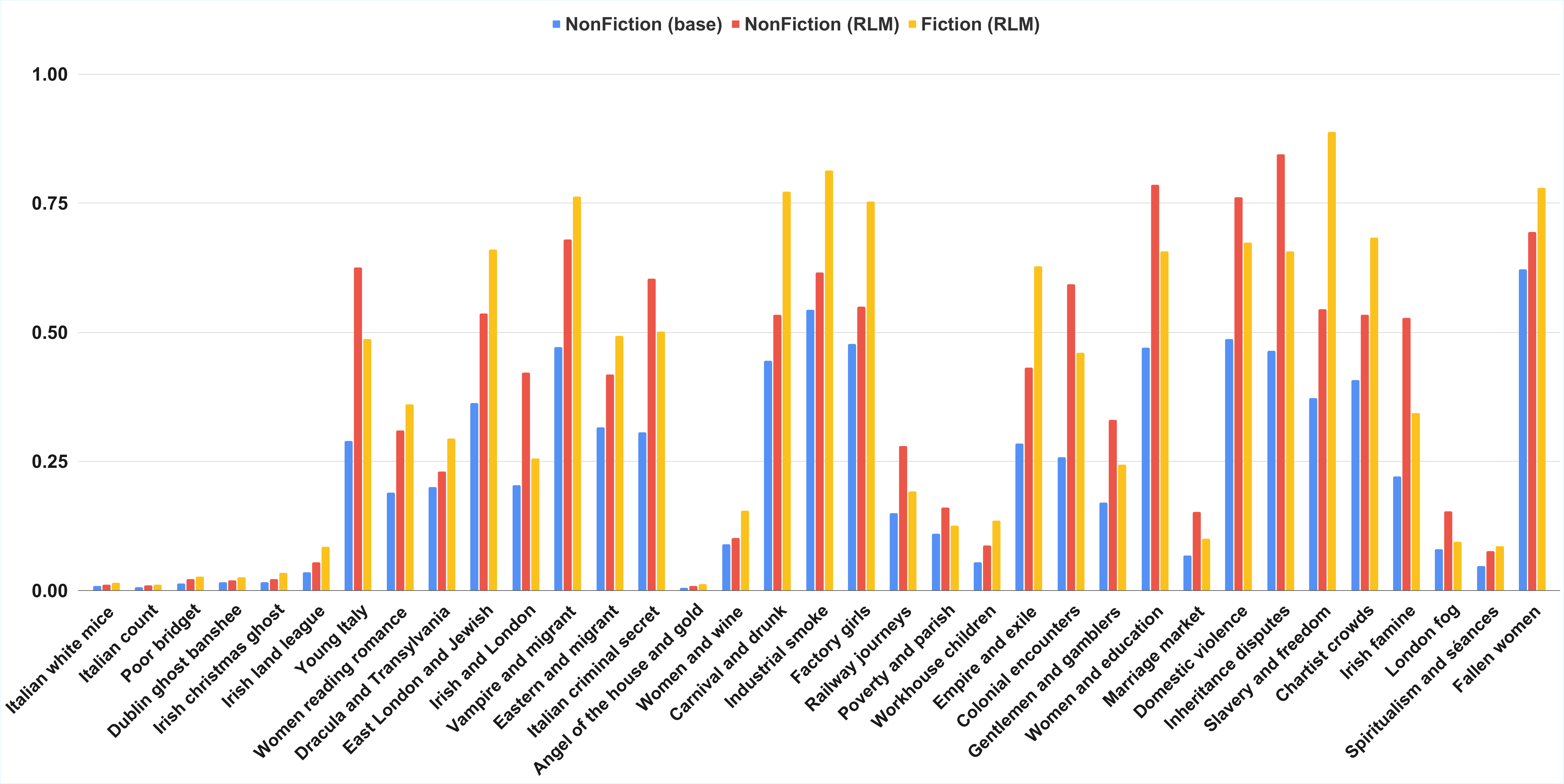}
% \vskip -0.1em
\caption{Comparisons of per query performance, measured by Average Precision (AP), among the three models from Table \ref{tab:result}: \nfbase, \nfrlm~and~\frlm.}
\label{fig:per_query_ap}
\end{figure*}

\begin{table}[!t]
\centering
\caption{
\small
Feedback terms with the respective term weights in the brackets, estimated via~\frlm~(left column) and~\nfrlm~(right column). 
Common terms that are estimated from both of the collections are bold-faced.
}
\begin{adjustbox}{width=1\textwidth}
\begin{tabular}{p{1.6cm}|@{~~}p{6cm}@{~~~} p{6cm}}

\toprule

\multicolumn{3}{c}{\textbf{Feedback Terms}}\\

\hline

\textbf{Topic} & \textbf{\frlm} & \textbf{\nfrlm} \\

\midrule

\textbf{Carnival and drunk}
& \textbf{festival} (0.092), revelry	(0.086), \textbf{mask} (0.081), laughter (0.078), \textbf{wine}	(0.075), \textbf{dance} (0.072), folly (0.068), \textbf{costume} (0.066), merry (0.063), jest (0.061), masquerade (0.058), music (0.056), theatre (0.054), spirits (0.051), debauchery (0.048), feast (0.046), jesters (0.044), song (0.042), crowd (0.04), laughter (0.038)
& consumption (0.084), intoxication (0.079), temperance (0.076), alcohol (0.073), \textbf{mask} (0.071), public	(0.068), disorder (0.066), \textbf{festival} (0.064), excess (0.061), vice (0.059), class (0.057), \textbf{wine} (0.055), abstinence (0.054), crime (0.052), drink (0.05), \textbf{costume}	(0.048), moral (0.047), labour (0.046), legislation (0.045), \textbf{dance} (0.044)
\\

\midrule

\textbf{Italian Criminal Secret Society}
& secracy (0.091), dagger (0.086), \textbf{republican} (0.081), vengeance (0.078), brotherhood (0.075), assassin (0.071), conspiracy (0.068), \textbf{political} (0.066), masked (0.063), \textbf{revolution}	(0.061), \textbf{Mazzini}	(0.059), nocturnal (0.056), avenger (0.053), dungeon	(0.051), initiation	(0.049), intrigue (0.046), \textbf{revolutionaries}	(0.044), monastery (0.041), \textbf{Carbonari} (0.038), \textbf{Rome} (0.036)
& \textbf{Carbonari} (0.084), \textbf{revolution} (0.081), \textbf{Mazzini} (0.079), \textbf{political} (0.077), society (0.074), secret (0.072), \textbf{republican} (0.07), organization	(0.068), Italy (0.066), uprising (0.065), \textbf{revolutionaries} (0.064), \textbf{Rome} (0.063), radical (0.061), association (0.06), reform (0.058), nationalist (0.057), liberation (0.055), Europe (0.053), revolutionist (0.052), unity (0.05) 
\\

\bottomrule

\end{tabular}
\label{tab:feedback}

\end{adjustbox}
\end{table}

\begin{table}[!t]
\centering
\small
\caption{
  \hlt{Topics that are predominantly fictional in 19th-century British writings, showing lowest Average Precision (AP) in non-fiction texts (see Figure \ref{fig:per_query_ap}).
  `Fictional Role' summarizes how each topic functions narratively or symbolically within literary texts; `Non-fiction Presence' indicates whether analogous discourse appears in the non-literary cases, highlighting the scarcity of such connections; `Explanation' contextualizes why the occurrence of each topic is largely confined to fictional content.}
}
\label{tab:explanation}
\begin{adjustbox}{width=1\textwidth}
\begin{tabular}{p{2.5cm}@{~~}|@{~~}p{4cm}@{~~~} p{4cm}@{~~~} p{4cm}} 

\toprule

\textbf{Topic} 
& \textbf{Fictional Role}
& \textbf{Non-fiction Presence}
& \textbf{Explanation} \\

\hline

\textbf{Italian white mice} 
& Comic or picturesque detail used in fiction to suggest urban curiosity, itinerant entertainers, or eccentric foreigners. 
& Almost none; performing animals mentioned only incidentally in travel writing or anecdotes. 
& Serves as whimsical exotic color in fiction, not as a subject of serious commentary or debate.\\

\textbf{Italian count} 
& A stock Gothic and sensation figure symbolizing continental intrigue, aristocratic decadence, or seductive danger.
& Rarely discussed in factual discourse; foreign nobility mentioned only incidentally in the press.
& Embodies the stereotype of the threatening or alluring foreigner; a purely narrative construct.
\\

\textbf{Poor bridget} 
& Represents the Irish maidservant in sentimental or comic fiction, often the object of sympathy or ridicule.
& Irish domestic labour discussed sociologically, but not under this name.
& The phrase functions as a literary shorthand for the Irish servant stereotype rather than a social analysis.
\\

\textbf{Dublin ghost/banshee} 
& Draws on Irish folklore and Gothic traditions to evoke ancestral haunting and national identity.
& Antiquarian and folklore studies mention banshees, but outside mainstream social discourse.
& Serves as a symbolic or supernatural expression of Irishness more than a factual or political topic.
\\

\textbf{Irish Christmas ghost} 
& A festive Gothic subgenre mixing sentimentality and the supernatural within an Irish setting.
& Limited to literary reviews and folklore anecdotes.
& A product of the Victorian publishing market and seasonal taste, not a subject of non-fictional debate.
\\

\textbf{Angels of the house and gold fields} 
& Reimagines the domestic ideal in colonial or emigration fiction, testing female virtue abroad.
& Women’s emigration discussed in reform writing, but without this symbolic framing.
& The specific pairing of domestic purity and frontier labour is a narrative trope rather than a documentary theme.
\\

\bottomrule
\end{tabular}
\end{adjustbox}
\end{table}

\subsection{Discussion}

Figure~\ref{fig:per_query_ap} depicts the Average Precision (AP) for each of the $35$ topics across the three retrieval settings, i.e. \nfbase, \nfrlm~and \frlm. 
It is observed from the bar plots that \nfrlm~(red bars in Figure~\ref{fig:per_query_ap}) tends to outperform the \nfbase~(blue bars) across most topical and socio-political themes (e.g. `Irish Land League', `Factory girls', `Slavery and freedom'), confirming that relevance modeling improves ranking effectiveness in expository domains.
In general, the BM25 baseline (i.e. \nfbase) remains relatively flat, indicating consistent lexical matching performance but missing domain-specific contextual enrichment provided by RLM-based expansion.

On the other hand, \frlm~(yellow bars) fluctuates more, performing well on some narrative or symbolic topics (e.g., `Dracula and Transylvania', `Vampire and migrant', `Spiritualism and séances'), however, underperforming on factually grounded queries (e.g., `Irish land league', `Industrial smoke', `Workhouse children').
This reflects that fiction-based feedback captures thematic associations and metaphorical language, which may not generalize effectively to expository non-fiction retrieval.

The advantage of \nfrlm~is pronounced for topics related to social reform, industrialization, and urban life (e.g. `Factory girls', `Chartist crowds', `Irish famine', `London fog').
Conversely, \frlm~excels for supernatural, emotional, or moral queries (e.g. `Irish Christmas ghost', `Fallen women', `Dracula and Transylvania'), supporting our hypothesis around transfer from fiction to non-fiction when working with interpretive or culturally encoded topics.

The feedback terms estimated from both \nfrlm~and~\frlm for two sample queries (Table~\ref{tab:feedback}) offer a closer look at how, despite shared thematic vocabularies, fiction and non-fiction diverge in their term-weight hierarchies by genre.
Table~\ref{tab:feedback} presents the feedback terms and their weights estimated by RLM for the queries `Carnival and drunk' (higher AP with \frlm) and `Italian criminal secret society' (higher AP with \nfrlm) in the fiction and non-fiction corpora. 
Higher weights indicate stronger relevance to the expanded query. 
While fiction terms emphasize imagery and emotion (festival, revelry, dagger, vengeance), non-fiction terms highlight social or political discourse (temperance, legislation, reform, organization). 
Bold-faced overlaps (wine, mask, costume, Carbonari, Mazzini, revolution) reveal shared vocabularies whose meanings shift across genres -- from aesthetic and sensational in fiction to moralistic or documentary in non-fiction. 
The differing weight patterns reflect how genre shapes the lexical and conceptual framing of similar cultural topics.

% On the other hand,
% \frlm~(blue bar in Figure~\ref{fig:per_query_ap}) outperforms other baselines for literary or symbolic topics (e.g. `Italian count', `Poor bridget', `Irish Christmas ghost'), suggesting that the model captures stylistic and narrative cues of fictions.
% \nfrlm~(red bar in Figure~\ref{fig:per_query_ap}), on the other hand, tends to outperform the \nfbase~across most topical and socio-political themes (e.g. `Irish Land League', `Factory girls', `Slavery and freedom'), confirming that relevance modeling improves ranking effectiveness in expository domains.
% The \nfbase~model consistently underperforms, indicating that without relevance feedback or contextual adaptation, precision at higher ranks drops sharply.
% % 
% Topics like `Vampire and migrant', `Fallen women', `Women and education' show higher AP scores in both fiction and non-fiction context (see the blue and red bars of the corresponding topics in Figure~\ref{fig:per_query_ap}), suggesting cross-genre topical coherence, i.e. these topics appear dominantly in both literary and factual texts.

The six topics having the least average precision, across retrieval settings, can be observed from Figure~\ref{fig:per_query_ap} -- `Italian white mice', ` Italian count', `Poor Bridget', `Dublin ghost/banshee', `Irish Christmas ghost', and `Angel of the house and gold fields'. 
These examples illustrate the retrieval difficulties introduced by figurative and narrative motifs.
The topics are likely to be highly symbolic and lexically unstable, often appearing in fiction through varied phrasing and context-specific expression (illustrated in Table~\ref{tab:explanation}). 
Their low document frequency and semantic ambiguity reduce the effectiveness of term-based relevance modeling, while their weak presence in non-fiction further limits cross-domain retrieval. 
The resulting low AP values across all models indicate that such motifs function primarily as imaginative or cultural constructs rather than consistently articulated subjects in textual discourse. 
This outcome demonstrates the limits of lexical retrieval models when applied to metaphorical or archetypal queries, and highlights the genre dependence of thematic representation in 19th-century British writing.

% Table \ref{tab:explanation} identifies six topics from the broader dataset that occur mainly within fictional rather than non-fictional discourse in 19th-century British writing. 
% The table illustrates how topic modeling or keyword-based retrieval might capture semantically rich but contextually limited motifs. 
% The `Fictional Role' column summarizes how each topic functions narratively or symbolically within literary texts. 
% The `Non-fiction Presence' column indicates whether analogous discourse appears in essays, journalism, or reform literature, highlighting the scarcity of such connections. 
% On the other hand, the `Explanation' column contextualizes why the occurance of each topic is largely confined to imaginative representation, helping to distinguish lexical presence from cultural salience. 
% Together, these columns provide interpretive grounding for computational findings, clarifying that not all co-occurrences imply cross-genre significance, which is an essential distinction in corpus-based cultural research.

\section{Conclusion and Future Work}

This study proposes a framework for cross-genre knowledge transfer in historical information retrieval, using the British Library 19th-century digital collection as a benchmark to examine links between fiction and non-fiction retrieval. 
Leveraging narrative and semantic patterns from fiction yielded measurable gains in non-fiction performance, especially in MAP and recall, showing that narrative richness can enhance conceptual breadth and contextual understanding. 
However, the findings also reveal an important trade-off. Specifically, we observed that while fiction contributes valuable contextual diversity, excessive genre mixing may reduce topical precision.

% This study introduced a novel framework for cross-genre knowledge transfer in historical information retrieval, using the British Library 19th-century digital collection as a benchmark for exploring the relationship between fiction and non-fiction retrieval.
% By leveraging narrative and semantic patterns from fiction, we demonstrated measurable improvements in non-fiction retrieval performance, particularly in MAP and recall, highlighting how narrative richness can broaden conceptual coverage and support deeper contextual understanding.
% While these results suggest that fiction-derived term distributions can serve as a valuable resource for enhancing retrieval in historical domains containing factual information, %promoting interpretability and inclusivity in cultural heritage search systems.
% we have observed a trade-off between semantic richness and topical precision.
% Although fiction can introduce useful contextual diversity, uncontrolled genre mixing may reduce retrieval focus.

While the results of cross-genre transfer are promising, further research is needed to verify whether similar effects extend to other centuries, languages, topics, and literary traditions. 
Such comparisons would help determine whether the observed improvements are broadly generalizable or specific to 19th-century stylistic and linguistic conventions. 
The current framework also does not fully address the expressive and structural differences between fiction and non-fiction texts. 
Future studies could investigate whether performance gains arise primarily from the richer semantic associations of fictional novels or from stylistic affinities between genres. 
Since combining materials from distinct genres can sometimes blur topical boundaries, introducing genre-aware weighting or adaptive fusion methods may help preserve the imaginative diversity of fiction while maintaining the thematic precision required for effective non-fiction retrieval.

% While the results of cross-genre transfer are promising, there are still several areas to explore further. It will be important to test whether similar effects hold for other centuries, languages, or genres. This will help to determine whether the observed improvements are general or specific to a certain period and style of writing. Further, the current setup does not account for the inherent diversity between fictional and non-fiction documents. Fiction often differs from non-fiction not only in content but also in how ideas are expressed. Future work could examine whether the benefits we observe come mainly from richer semantics or from stylistic similarities between genres. We have seen that combining information from different genres into a single model can sometimes blur topical focus. Hence, introducing genre-aware weighting could help balance the contribution of each type of text, preserving the diversity of fiction while maintaining the precision of non-fiction retrieval.

\subsubsection*{\textbf{Acknowledgement.}}
% \small 
This publication is part of a project that has received funding from (i) the European Research Council (ERC) under the Horizon 2020
research and innovation program (Grant agreement No. 884951); (ii) Research Ireland to the Insight Centre for Data
Analytics under grant No 12/RC/2289\_P2. 
We would also like to sincerely thank the annotators from the UCD School of English, Drama and Film for their invaluable efforts in annotating the data. Their careful, consistent, and time-intensive work was essential to the quality and reliability of this study.
% We greatly appreciate their dedication and contribution to this research.
% 

\subsubsection*{\textbf{Disclosure of Interests.}}
The authors declare no competing interests that influenced the research, authorship, or publication of this article.
The first two authors acknowledge support from a lab visit grant from GESIS -- Leibniz Institute for the Social Sciences, Cologne, Germany, which facilitated a research visit during which the initial planning and conceptual discussions for this work took place. 
% GESIS provided funding for travel through its visiting researchers program. 
However, the support received from GESIS did not influence the research design, data collection, analysis, interpretation, or writing of the manuscript.

\bibliographystyle{splncs04}
\bibliography{refs}

@article{MARSH2003519,
title = {Learning facts from fiction},
journal = {Journal of Memory and Language},
volume = {49},
number = {4},
pages = {519-536},
year = {2003},
author = {Elizabeth J Marsh and Michelle L Meade and Henry L {Roediger III}},
}

@article{Azagra_Caro_2018,
   title={‘Getting out of the closet’: scientific authorship of literary fiction and knowledge transfer},
   volume={45},
   number={1},
   journal={The Journal of Technology Transfer},
   publisher={Springer Science and Business Media LLC},
   author={Azagra-Caro, Joaquín M. and Fernández-Mesa, Anabel and Robinson-García, Nicolás},
   year={2018},
   month=jun, pages={56–85} }

@inproceedings{jurczyk-choi-2017-cross,
    title = "Cross-genre Document Retrieval: Matching between Conversational and Formal Writings",
    author = "Jurczyk, Tomasz  and
      Choi, Jinho D.",
    booktitle = "Proceedings of the First Workshop on Building Linguistically Generalizable {NLP} Systems",
    month = sep,
    year = "2017",
    address = "Copenhagen, Denmark",
    publisher = "Association for Computational Linguistics",
    pages = "48--53",
}

@inproceedings{mostafazadeh-etal-2016-corpus,
    title = "A Corpus and Cloze Evaluation for Deeper Understanding of Commonsense Stories",
    author = "Mostafazadeh, Nasrin  and
      Chambers, Nathanael  and
      He, Xiaodong  and
      Parikh, Devi  and
      Batra, Dhruv  and
      Vanderwende, Lucy  and
      Kohli, Pushmeet  and
      Allen, James",
    booktitle = "Proceedings of the 2016 Conference of the North {A}merican Chapter of the Association for Computational Linguistics: Human Language Technologies",
    month = jun,
    year = "2016",
    address = "San Diego, California",
    publisher = "Association for Computational Linguistics",
    pages = "839--849"
}

@misc{bert,
      title={BERT: Pre-training of Deep Bidirectional Transformers for Language Understanding}, 
      author={Jacob Devlin and Ming-Wei Chang and Kenton Lee and Kristina Toutanova},
      year={2019},
}

@misc{roberta,
      title={RoBERTa: A Robustly Optimized BERT Pretraining Approach}, 
      author={Yinhan Liu and Myle Ott and Naman Goyal and Jingfei Du and Mandar Joshi and Danqi Chen and Omer Levy and Mike Lewis and Luke Zettlemoyer and Veselin Stoyanov},
      year={2019},
      eprint={1907.11692},
      archivePrefix={arXiv},
      primaryClass={cs.CL},
      url={https://arxiv.org/abs/1907.11692}, 
}

@misc{bookcorpus,
  title = {{Book Corpus}},
  howpublished = {\url{https://en.wikipedia.org/wiki/BookCorpus}},
  note = {Accessed: 2015-10-28}
}

@inbook{datta2025jcdl,
author = {Datta, Suchana and Roy, Dwaipayan and Greene, Derek and Meaney, Gerardine},
title = {Unveiling Temporal Trends in 19th Century Literature: An Information Retrieval Approach},
year = {2025},
isbn = {9798400710933},
publisher = {Association for Computing Machinery},
address = {New York, NY, USA},
url = {https://doi.org/10.1145/3677389.3702593},
booktitle = {Proceedings of the 24th ACM/IEEE Joint Conference on Digital Libraries},
articleno = {22},
numpages = {5}
}

@inproceedings{datta2025ecir,
author = {Datta, Suchana and Roy, Dwaipayan and Greene, Derek and Meaney, Gerardine},
title = {Tales and Truths: Exploring the Linguistic Journey of 19th Century Literature and Non-fiction},
year = {2025},
isbn = {978-3-031-88716-1},
publisher = {Springer-Verlag},
address = {Berlin, Heidelberg},
doi = {10.1007/978-3-031-88717-8_19},
booktitle = {Advances in Information Retrieval: 47th European Conference on Information Retrieval, ECIR 2025, Lucca, Italy, April 6–10, 2025, Proceedings, Part IV},
pages = {252–266},
numpages = {15},
location = {Lucca, Italy}
}

@InProceedings{bm25,
author="Robertson, S. E.
and Walker, S.",
title="Some Simple Effective Approximations to the 2-Poisson Model for Probabilistic Weighted Retrieval",
booktitle="SIGIR '94",
year="1994",
publisher="Springer London",
address="London",
pages="232--241",
}

@article{bm25_beyond,
author = {Robertson, Stephen and Zaragoza, Hugo},
title = {The Probabilistic Relevance Framework: {BM25} and Beyond},
year = {2009},
issue_date = {April 2009},
publisher = {Now Publishers Inc.},
address = {Hanover, MA, USA},
volume = {3},
number = {4},
journal = {Foundations and Trends in Information Retrieval},
month = {apr},
pages = {333–389},
numpages = {57}
}

@inproceedings{DBLP:conf/trec/JaleelACDLLSW04,
  author       = {Nasreen Abdul Jaleel and
                  James Allan and
                  W. Bruce Croft and
                  Fernando Diaz and
                  Leah S. Larkey and
                  Xiaoyan Li and
                  Mark D. Smucker and
                  Courtney Wade},
  title        = {UMass at {TREC} 2004: Novelty and {HARD}},
  booktitle    = {Proceedings of the Thirteenth Text REtrieval Conference},
  series       = {{NIST} Special Publication},
  volume       = {500-261},
  publisher    = {National Institute of Standards and Technology {(NIST)}},
  year         = {2004},
}

@inproceedings{10.1145/383952.383972,
author = {Lavrenko, Victor and Croft, W. Bruce},
title = {Relevance based language models},
year = {2001},
publisher = {Association for Computing Machinery},
address = {New York, NY, USA},
booktitle = {Proceedings of the 24th Annual International ACM SIGIR Conference on Research and Development in Information Retrieval},
pages = {120–127},
numpages = {8},
location = {New Orleans, Louisiana, USA},
series = {SIGIR '01}
}

@Article{chi2024diachronic,
AUTHOR = {Chi, Yang and Giunchiglia, Fausto and Xu, Hao},
TITLE = {Diachronic Semantic Tracking for Chinese Words and Morphemes over Centuries},
JOURNAL = {Electronics},
VOLUME = {13},
YEAR = {2024},
NUMBER = {9},
ARTICLE-NUMBER = {1728},
ISSN = {2079-9292},
DOI = {10.3390/electronics13091728}
}

@inproceedings{hamilton2016diachronic,
    title = "Diachronic Word Embeddings Reveal Statistical Laws of Semantic Change",
    author = "Hamilton, William L.  and
      Leskovec, Jure  and
      Jurafsky, Dan",
    editor = "Erk, Katrin  and
      Smith, Noah A.",
    booktitle = "Proceedings of the 54th Annual Meeting of the Association for Computational Linguistics (Volume 1: Long Papers)",
    month = aug,
    year = "2016",
    address = "Berlin, Germany",
    publisher = "Association for Computational Linguistics",
    doi = "10.18653/v1/P16-1141",
    pages = "1489--1501"
}

@inproceedings{lv2010prlm,
author = {Lv, Yuanhua and Zhai, ChengXiang},
title = {Positional relevance model for pseudo-relevance feedback},
year = {2010},
isbn = {9781450301534},
publisher = {Association for Computing Machinery},
address = {New York, NY, USA},
doi = {10.1145/1835449.1835546},
booktitle = {Proceedings of the 33rd International ACM SIGIR Conference on Research and Development in Information Retrieval},
pages = {579–586},
numpages = {8},
location = {Geneva, Switzerland},
series = {SIGIR '10}
}

@inproceedings{diaz2016local,
    title = "Query Expansion with Locally-Trained Word Embeddings",
    author = "Diaz, Fernando  and
      Mitra, Bhaskar  and
      Craswell, Nick",
    editor = "Erk, Katrin  and
      Smith, Noah A.",
    booktitle = "Proceedings of the 54th Annual Meeting of the Association for Computational Linguistics (Volume 1: Long Papers)",
    month = aug,
    year = "2016",
    address = "Berlin, Germany",
    publisher = "Association for Computational Linguistics",
    doi = "10.18653/v1/P16-1035",
    pages = "367--377"
}

@article{mar2018stories,
  title     = {Stories and the Promotion of Social Cognition},
  author    = {Mar, Raymond A.},
  journal   = {Current Directions in Psychological Science},
  volume    = {27},
  number    = {4},
  pages     = {257--262},
  year      = {2018},
  publisher = {SAGE Publications},
  doi       = {10.1177/0963721417749654}
}

@article{jacobs2015neurocognitive,
  title     = {Towards a Neurocognitive Poetics Model of Literary Reading},
  author    = {Jacobs, Arthur M.},
  journal   = {Cognitive Critique},
  volume    = {11},
  pages     = {1--24},
  year      = {2015}
}

@article{bendavid2010theory,
  title     = {A Theory of Learning from Different Domains},
  author    = {Ben-David, Shai and Blitzer, John and Crammer, Koby and Pereira, Fernando},
  journal   = {Machine Learning},
  volume    = {79},
  number    = {1--2},
  pages     = {151--175},
  year      = {2010},
  publisher = {Springer},
  doi       = {10.1007/s10994-009-5152-4}
}

@inproceedings{zamani2018neural,
  title     = {From Neural Re-Ranking to Neural Ranking: Learning a Sparse Representation for Information Retrieval},
  author    = {Zamani, Hamed and Dehghani, Mostafa and Croft, W. Bruce and Learned-Miller, Erik and Kamps, Jaap},
  booktitle = {Proceedings of the 27th ACM International Conference on Information and Knowledge Management (CIKM 2018)},
  pages     = {497--506},
  year      = {2018},
  doi       = {10.1145/3269206.3271773}
}

@inproceedings{wang2018glue,
    title = "{GLUE}: A Multi-Task Benchmark and Analysis Platform for Natural Language Understanding",
    author = "Wang, Alex  and
      Singh, Amanpreet  and
      Michael, Julian  and
      Hill, Felix  and
      Levy, Omer  and
      Bowman, Samuel",
    editor = "Linzen, Tal  and
      Chrupa{\l}a, Grzegorz  and
      Alishahi, Afra",
    booktitle = "Proceedings of the 2018 {EMNLP} Workshop {B}lackbox{NLP}: Analyzing and Interpreting Neural Networks for {NLP}",
    month = nov,
    year = "2018",
    address = "Brussels, Belgium",
    publisher = "Association for Computational Linguistics",
    doi = "10.18653/v1/W18-5446",
    pages = "353--355",
}

@inproceedings{rajpurkar2018squad,
    title = "Know What You Don{'}t Know: Unanswerable Questions for {SQ}u{AD}",
    author = "Rajpurkar, Pranav  and
      Jia, Robin  and
      Liang, Percy",
    editor = "Gurevych, Iryna  and
      Miyao, Yusuke",
    booktitle = "Proceedings of the 56th Annual Meeting of the Association for Computational Linguistics (Volume 2: Short Papers)",
    month = jul,
    year = "2018",
    address = "Melbourne, Australia",
    publisher = "Association for Computational Linguistics",
    doi = "10.18653/v1/P18-2124",
    pages = "784--789",
}

@inbook{cleverdon,
author = {Cleverdon, Cyril},
title = {The Cranfield tests on index language devices},
year = {1997},
isbn = {1558604545},
publisher = {Morgan Kaufmann Publishers Inc.},
address = {San Francisco, CA, USA},
booktitle = {Readings in Information Retrieval},
pages = {47–59},
numpages = {13}
}

@inproceedings{ance,
  author       = {Lee Xiong and
                  Chenyan Xiong and
                  Ye Li and
                  Kwok{-}Fung Tang and
                  Jialin Liu and
                  Paul N. Bennett and
                  Junaid Ahmed and
                  Arnold Overwijk},
  title        = {Approximate Nearest Neighbor Negative Contrastive Learning for Dense
                  Text Retrieval},
  booktitle    = {9th International Conference on Learning Representations, {ICLR} 2021,
                  Virtual Event, Austria, May 3-7, 2021},
  publisher    = {OpenReview.net},
  year         = {2021},
  url          = {https://openreview.net/forum?id=zeFrfgyZln},
}

@misc{meng2024aug,
      title={AugTriever: Unsupervised Dense Retrieval and Domain Adaptation by Scalable Data Augmentation}, 
      author={Rui Meng and Ye Liu and Semih Yavuz and Divyansh Agarwal and Lifu Tu and Ning Yu and Jianguo Zhang and Meghana Bhat and Yingbo Zhou},
      year={2024},
      eprint={2212.08841},
      archivePrefix={arXiv},
      primaryClass={cs.CL},
      url={https://arxiv.org/abs/2212.08841}, 
}

@inproceedings{santhanam2022colbertv2,
    title = "{C}ol{BERT}v2: Effective and Efficient Retrieval via Lightweight Late Interaction",
    author = "Santhanam, Keshav  and
      Khattab, Omar  and
      Saad-Falcon, Jon  and
      Potts, Christopher  and
      Zaharia, Matei",
    editor = "Carpuat, Marine  and
      de Marneffe, Marie-Catherine  and
      Meza Ruiz, Ivan Vladimir",
    booktitle = "Proceedings of the 2022 Conference of the North American Chapter of the Association for Computational Linguistics: Human Language Technologies",
    month = jul,
    year = "2022",
    address = "Seattle, United States",
    publisher = "Association for Computational Linguistics",
    doi = "10.18653/v1/2022.naacl-main.272",
    pages = "3715--3734",
}

@misc{nogueira2019passage,
      title={Passage Re-ranking with BERT}, 
      author={Rodrigo Nogueira and Kyunghyun Cho},
      year={2020},
      eprint={1901.04085},
      archivePrefix={arXiv},
      primaryClass={cs.IR},
      url={https://arxiv.org/abs/1901.04085}, 
}

@article{temporal,
    author = {Dhingra, Bhuwan and Cole, Jeremy R. and Eisenschlos, Julian Martin and Gillick, Daniel and Eisenstein, Jacob and Cohen, William W.},
    title = {Time-Aware Language Models as Temporal Knowledge Bases},
    journal = {Transactions of the Association for Computational Linguistics},
    volume = {10},
    pages = {257-273},
    year = {2022},
}

@ARTICLE{adaptation,
  author={Singhal, Peeyush and Walambe, Rahee and Ramanna, Sheela and Kotecha, Ketan},
  journal={IEEE Access}, 
  title={Domain Adaptation: Challenges, Methods, Datasets, and Applications}, 
  year={2023},
  volume={11},
  number={},
  pages={6973-7020},
  }

@inproceedings{vocab_shift,
author = {Kenter, Tom and Wevers, Melvin and Huijnen, Pim and de Rijke, Maarten},
title = {Ad Hoc Monitoring of Vocabulary Shifts over Time},
year = {2015},
publisher = {Association for Computing Machinery},
address = {New York, NY, USA},
booktitle = {Proceedings of the 24th ACM International on Conference on Information and Knowledge Management},
pages = {1191–1200},
numpages = {10},
location = {Melbourne, Australia},
series = {CIKM '15}
}

@article{carpineto2001info,
author = {Carpineto, Claudio and de Mori, Renato and Romano, Giovanni and Bigi, Brigitte},
title = {An information-theoretic approach to automatic query expansion},
year = {2001},
publisher = {Association for Computing Machinery},
address = {New York, NY, USA},
volume = {19},
number = {1},
issn = {1046-8188},
doi = {10.1145/366836.366860},
journal = {ACM Trans. Inf. Syst.},
month = jan,
pages = {1–27},
numpages = {27}
}

@article{transfer1,
 author = {Kai Mikkonen},
 journal = {Style},
 number = {4},
 pages = {291--312},
 publisher = {Penn State University Press},
 title = {Can Fiction Become Fact? The Fiction-to-Fact Transition in Recent Theories of Fiction},
 urldate = {2025-10-29},
 volume = {40},
 year = {2006}
}

@article{transfer2,
 author = {Debbie Stien and Penny L. Beed},
 journal = {The Reading Teacher},
 number = {6},
 pages = {510--518},
 publisher = {[Wiley, International Reading Association]},
 title = {Bridging the Gap between Fiction and Nonfiction in the Literature Circle Setting},
 urldate = {2025-10-29},
 volume = {57},
 year = {2004}
}

@article{valdez,
author = {Jessica R. Valdez },
title = {Victorian Studies, Literature, and the Global Nineteenth Century},
journal = {Global Nineteenth-Century Studies},
volume = {1},
number = {1},
pages = {37-42},
year = {2022},
}
\end{document}